\documentclass[10pt,journal]{IEEEtran}
\usepackage[utf8]{inputenc}
\usepackage{url}
\usepackage[table]{xcolor}
\usepackage{authblk}
\usepackage[numbers]{natbib}
\usepackage{subfig}
\usepackage{booktabs}
\usepackage{lscape} 
\usepackage{tabularx}
\usepackage{ragged2e}
\usepackage{caption}
\usepackage{amssymb}% http://ctan.org/pkg/amssymb
\usepackage{pifont}% http://ctan.org/pkg/pifont
\usepackage{textcomp}
\usepackage{multirow}
\usepackage[normalem]{ulem}
\usepackage[flushleft]{threeparttable}
\useunder{\uline}{\ul}{}
\usepackage{graphicx}
\usepackage[colorlinks=true,urlcolor= blue,linkcolor=black]{hyperref}%
\usepackage{bbding}

\definecolor{lightgray}{gray}{0.9}
\newcommand{\xmark}{\ding{55}}%

\newcolumntype{L}{>{\RaggedRight\arraybackslash}X}

\begin{document}

\title{Improving the generalizability of convolutional neural network-based segmentation on CMR images}

% Default path to search graphics
\author[1]{\IEEEauthorblockN{Chen Chen}$^{\textrm{(\Envelope)}}$~\thanks{chen.chen15@imperial.ac.uk}}
\author[2,3]{\IEEEauthorblockN {Wenjia Bai}}
\author[4]{\IEEEauthorblockN Rhodri H. Davies}
\author[4,5]{\IEEEauthorblockN Anish N. Bhuva}
\author[4,5]{\IEEEauthorblockN Charlotte Manisty}
\author[4,5]{\IEEEauthorblockN James C. Moon}

\author[5,6]{\IEEEauthorblockN Nay Aung}
\author[6]{\IEEEauthorblockN Aaron M.Lee}
\author[5,6]{\IEEEauthorblockN Mihir M. Sanghvi}
\author[5,6]{\IEEEauthorblockN Kenneth Fung}
\author[6]{\IEEEauthorblockN Jose Miguel Paiva}
\author[5,6]{\IEEEauthorblockN Steffen E. Petersen}
\author[7]{\IEEEauthorblockN Elena Lukaschuk}
\author[7]{\IEEEauthorblockN Stefan K. Piechnik}
\author[7]{\IEEEauthorblockN Stefan Neubauer}

\author[1]{\IEEEauthorblockN Daniel Rueckert}

\affil[1]{\footnotesize Biomedical Image Analysis Group, Department of Computing, Imperial College London, London, UK}
\affil[2]{\footnotesize Data Science Institute, Imperial College London, London, UK}
\affil[3]{\footnotesize Department of Medicine, Imperial College London, London, UK}
\affil[4]{\footnotesize Institutes for Cardiovascular Science, University College London, London, UK}

\affil[5]{\footnotesize Department of Cardiovascular Imaging, Barts Heart Centre, St Bartholomew's Hospital, London, UK}
\affil[6]{\footnotesize NIHR Biomedical Research Centre at Barts, Queen Mary University of London, London, UK
}
\affil[7]{\footnotesize NIHR BRC Oxford, Division of Cardiovascular Medicine, Radcliffe, Department of Medicine, University of Oxford, Oxford, UK.
}

\maketitle

\begin{abstract}
Convolutional neural network (CNN) based segmentation methods provide an efficient and automated way for clinicians to assess the structure and function of the heart in cardiac MR images. While CNNs can generally perform the segmentation tasks with high accuracy when training and test images come from the same domain (e.g. same scanner or site), their performance often degrades dramatically on images from different scanners or clinical sites. We propose a simple yet effective way for improving the network generalization ability by carefully designing data normalization and augmentation strategies to accommodate common scenarios in multi-site, multi-scanner clinical imaging data sets. We demonstrate that a neural network trained on a \emph{single-site single-scanner} dataset from the UK Biobank can be successfully applied to segmenting cardiac MR images across \emph{different sites} and \emph{different scanners} without substantial loss of accuracy. Specifically, the method was trained on a large set of 3,975 subjects from the UK Biobank. It was then \emph{directly} tested on 600 different subjects from the UK Biobank for intra-domain testing and two other sets for cross-domain testing: the ACDC dataset (100 subjects, 1 site, 2 scanners) and the BSCMR-AS dataset (599 subjects, 6 sites, 9 scanners). The proposed method produces promising segmentation results on the UK Biobank test set which are comparable to previously reported values in the literature, while also performing well on cross-domain test sets, achieving a mean Dice metric of 0.90 for the left ventricle, 0.81 for the myocardium and 0.82 for the right ventricle on the ACDC dataset; and 0.89 for the left ventricle, 0.83 for the myocardium on the BSCMR-AS dataset. The proposed method offers a potential solution to improve CNN-based model generalizability for the cross-scanner and cross-site cardiac MR image segmentation task. 
\end{abstract}
\vspace{0pt}

\section{Introduction}
\begin{figure*}[!ht]
\centerline{
\includegraphics[width=0.8\textwidth]{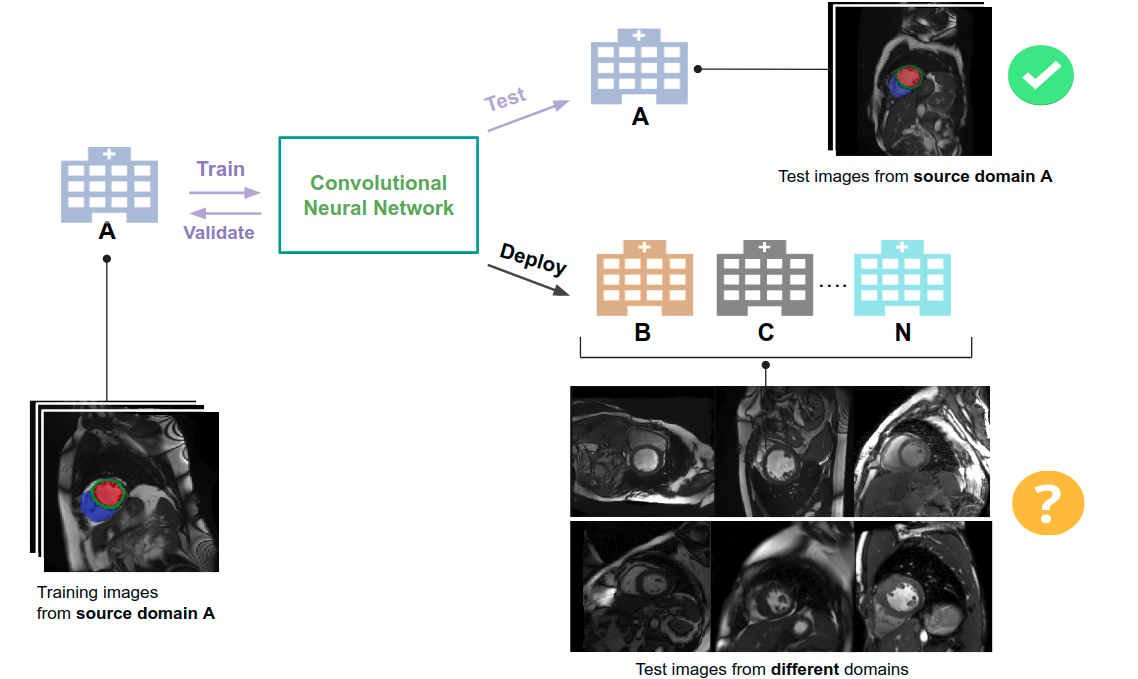}}
\caption{\textbf{Illustration of a cross-domain CMR image segmentation application}
A CNN model which has been trained using a dataset collected from site A (source domain) is deployed onto data from other sites (target domains) to segment the left ventricle, the myocardium and the right ventricle from CMR images. In general, the model can perform well on test images from the same domain. However, whether this model can generalise well onto other sites is unknown. For example, the training set may have limited pathological cases, which may cause the model not be able to generalise over subjects with heart conditions outside of the training set. In addition, images from different scanners may have different image appearance because of different imaging acquisition protocols. Both these differences pose challenges to applying a CNN-based cardiac segmentation model to everyday clinical practice.}
\label{Overview}
\end{figure*}

Automatic cardiac segmentation algorithms provide an efficient way for clinicians to assess the structure and function of the heart from cardiac magnetic resonance (CMR) images for the diagnosis and management of a wide range of abnormal heart conditions \cite{Bernard2018}. Recently, convolutional neural network (CNN)-based methods have become state-of-the-art techniques for automated cardiac image segmentation \cite{greenspan2016guest,Bernard2018}. However, related work \cite{Bai2018} has shown that the segmentation accuracy of a CNN may degrade if the network is directly applied to images collected from different sites or scanners. For instance, CMR images coming from different sites may comprise different population demographics in terms of cardiovascular diseases, resulting in the clinically appreciable difference not only in cardiac morphology but also in image quality (e.g. irregular heartbeat can affect image quality)~\cite{Medrano-Gracia2014,rajwani2016incremental,petitjean2011review}. Thus, a CNN learned from a limited dataset may not be able to generalise over subjects with heart conditions outside of the training set. In addition, images from different scanners using different acquisition protocols can exhibit differences in terms of noise levels, image contrast, and resolution~\cite{Rajiah, Alfudhili2016,gutberlet2006comprehensive}. All these differences pose challenges for deploying CNN-based image segmentation algorithms in real-world practice, as illustrated in Fig.~\ref{Overview}.

%% related work and clarify the gap
In general, a straightforward way to address this problem is to fine-tune a CNN learned from one dataset (source domain) with additional labelled data from another dataset (target domain). Nevertheless, collecting sufficient pixel-wise labelled medical data for every scenario can be difficult, since it requires domain-specific knowledge and intensive labour to perform manual annotation. To alleviate the labelling cost, unsupervised deep domain adaptation (UDDA) approaches have been proposed \cite{wang2018deep}. Compared to fine-tuning, UDDA does not require labelled data from the target domain. Instead, it only uses either feature-level information \cite{Sun2016,long2015learning,Hoffman2017} or image-level information \cite{Hoffman2017} to optimize the network performance on the target domain. However, these methods usually require hand-crafted hyper-parameter tuning for each scenario, which may be difficult to scale to highly heterogeneous datasets. Therefore, it is of great interest to explore how to learn a network that can be successfully applied to other datasets without the requirement of additional model tuning.

%% our work
In this paper, we investigate the possibility of building a generalisable model for cardiac MR image segmentation, given a training set from only one scanner in a single site. Instead of fine-tuning or adapting to get a new model for each particular scenario, our goal is to find a generalisable solution that can analyse `real-world' test images collected from multiple sites and scanners. These images consist of various pathology and cardiac morphology that may not be present in the training set, reflecting the complexity of a real-world clinical setting. To achieve this goal, we choose the U-Net~\cite{Ronneberger2015a} as the fundamental CNN architecture and apply it to segment the cardiac anatomy from CMR images (short-axis view), including the left ventricle (LV), the myocardium (MYO), and the right ventricle (RV). An image pre-processing pipeline is proposed to normalise images across sites before feeding them to the network in both training and testing stages. Data augmentation is employed during the training to improve the generalization ability of the network. Although there has been a number of works \cite{Baumgartner2017,Tran2016} which have already applied data normalization and data augmentation in their pipelines, these methods are particularly designed for one specific dataset and the importance of applying data augmentation for model generalization ability across datasets is less explored. Here we demonstrate that the proposed data normalization and augmentation strategies can greatly improve the model performance in the cross-dataset setting (section~\ref{SECTION: data aug and norm impact}). 
The main contributions of the work are as follows:
\begin{itemize}
\item To the best of our knowledge, this is the first work to explore the generalizability of CNN-based methods for cardiac MR image multi-structure segmentation, where the training data is collected from a \textbf{single scanner}, but the test data comes from \textbf{multiple scanners} and \textbf{multiple sites}.
\item The proposed pipeline which employs data normalization and data augmentation (section~\ref{SECTION: data preprocessing}) is simple yet efficient and can be applied to training and testing of many state-of-the-art CNN architectures to improve the model segmentation accuracy across domains without necessarily sacrificing the accuracy in the original domain. Experiment results show that the proposed segmentation method is capable of segmenting multi-scanner, multi-vendor and multi-site datasets (section~\ref{SECTION: different scanners} and \ref{SECTION: different sites}).
\item Our work reveals that significant cardiac shape deformation caused by cardiac pathologies (section~\ref{SECTION: pathological effects}), low image quality (section~\ref{SECTION: pathological effects}), and inconsistent labelling protocols among different datasets (section~\ref{limitation}) are still major challenges for generalising deep learning-based cardiac image segmentation algorithms to images collected across different sites, which deserve further study.
\end{itemize}

\section{Related Work}
%% related literature 
There have been a great number of works which develop sophisticated deep learning approaches to perform CMR image segmentation tasks on a specific dataset~\cite{Tran2016, Bernard2018, Baumgartner2017,Bai2018}. While these models can achieve overall high accuracy over the samples from the same dataset, only a few have been explored in cross-dataset settings. Table~\ref{related literatures} shows a list of related works that demonstrate the segmentation performance of their proposed method by first training a model from one set (source domain) and then testing it on other datasets (target domain). However, these approaches requires re-training or fine-tuning to improve the performance on the target domain in a fully supervised fashion. To the best of our knowledge, there are few studies reported in the literature which investigate the generalization ability of the cardiac segmentation networks that can directly work across multiple sites.
\begin{table}[!ht]
\centering
\caption{\textbf{Related work that applies CNN-based CMR image segmentation models across multiple datasets}.}
\label{related literatures}
\resizebox{0.5\textwidth}{!}{%
\begin{tabular}{@{}ccccc@{}}
\toprule
\multirow{2}{*}{\textbf{Methods}} & \multirow{2}{*}{\begin{tabular}[c]{@{}c@{}}\textbf{Target domain} \\$\neq$  \textbf{Source domain}\end{tabular}} & \multirow{2}{*}{\textbf{Finetuning}} & \multirow{2}{*}{\textbf{Test on}} & \multirow{2}{*}{\begin{tabular}[c]{@{}c@{}}\textbf{Test set(s)} \\ \textbf{size}\end{tabular}} \\
 &  &  &  &  \\ \midrule
Tran, 2016~\cite{Tran2016} & Yes & Yes & \begin{tabular}[c]{@{}c@{}}LV/MYO/RV\\ separately\end{tabular} & \textless{}200 \\
Bai, 2018~\cite{Bai2018} & Yes & Yes & LV+MYO+RV & \textless{}100 \\
% Tao, 2018 \cite{Tao2018} (*) & No & No & MYO & \textless 200 \\
Khened, 2019 \cite{khened2018fully} & Yes & No & MYO & \textless 200 \\
Our work & Yes & No & LV+MYO+RV & 699 \\ \bottomrule
\end{tabular}
}
\end{table}

One work~\cite{Tao2018} in this line of research has been recently presented, which integrates training samples from multiple sites and multiple vendors \cite{Tao2018} to improve segmentation performance across sites. Their results show that the best segmentation performance on their multi-scanner test set was achieved when the data used for training and testing are from the same scanners. Nevertheless, their solution requires collecting annotated data from multiple vendors and sites. For deployment, this may not always be practical because of the high data collection and labelling costs as well as data privacy issues.  

Another direction to improve model generalization is to optimize the CNN architecture. In \cite{khened2018fully}, the authors proposed a novel network structure with residual connections to improve the network generalizability. They proposed that networks with a large number of parameters may easily suffer from over-fitting problem with limited data \cite{khened2018fully}. They showed that their light-weight network trained on a limited dataset outperformed the U-Net~\cite{Ronneberger2015a}, achieving higher accuracy on LV, myocardium, and RV. Moreover, model generalization was demonstrated by testing it (without any re-training or fine-tuning) on the LV-2011 dataset~\cite{Suinesiaputra2014}. As a result, this model produced comparable results to those from a network that had been trained on the LV-2011, achieving a high mean Dice score for the myocardium (0.84). However, because of the lack of RV labels in their test set, their network's performance on the RV is unclear. Segmenting the RV is considered to be harder than the LV and the myocardium because the RV has a more complex shape with higher variability across individuals, and its walls are thinner, making it harder to delineate from its surroundings. 

In this study, we evaluate the generalizability of the proposed method not only on the cardiac left ventricle segmentation but also on the right ventricle segmentation. Different from~\cite{Tao2018,khened2018fully}, the proposed method demonstrates model generalizability in a more challenging but realistic setting: our training data was collected from only one scanner (most of them are healthy subjects) while test data was collected from various unseen sites and scanners, which covers a wide range of pathologies, reflecting the spectrum of clinical practice.

\section{Materials and Methodology}
\subsection{Data}
\label{SECTION: dataset}
Three datasets are used in this study and the general descriptions of them are summarised in Table~\ref{table_for_the_threesdatasets_description}. 

\begin{table*}[!ht]
\caption{\textbf{General descriptions of the three datasets}}
\label{table_for_the_threesdatasets_description}
\resizebox{\textwidth}{!}{%
\begin{tabular}{@{}lllcll@{}}
\toprule
\multicolumn{1}{c}{\textbf{Name}} &
\multicolumn{1}{c}{\textbf{\begin{tabular}[c]{@{}c@{}}Number of \\ Subjects\end{tabular}}} & \multicolumn{1}{c}{\textbf{Cohort}} & \textbf{Sites} & \multicolumn{1}{c}{\textbf{Scanners}} & \textbf{\begin{tabular}[c]{@{}l@{}} Image Spatial \\ Resolution\end{tabular}} \\ \midrule
UKBB & 4875 & \begin{tabular}[c]{@{}l@{}} General population\\ \end{tabular} & 1 & 1.5 T, Aera, Siemens (100\%) & \begin{tabular}[c]{@{}l@{}}in-plane resolution: \\ 1.8 $mm^2$ /pixel;\\ slice thickness: \\ 8 mm\end{tabular} \\
%\rowcolor{lightgray}
\hline
ACDC & 100 & \begin{tabular}[c]{@{}l@{}}Without cardiac disease (20\%); \\ Dilated cardiomyopathy (20\%); \\ Hypertrophic cardiomyopathy (20\%); \\ Myocardial infarction with altered left \\ ventricular ejection (20\%); \\ Abnormal right ventricle (20\%)\end{tabular} & 1 & \begin{tabular}[c]{@{}l@{}}1.5 T, Area, Siemens (67\%)\\ 3 T, Trio Tim, Siemens (33\%)\end{tabular} & \begin{tabular}[c]{@{}l@{}}in-plane resolution: \\ 1.34 - 1.68 $mm^2$ /pixel; \\ slice thickness: \\ 5 -10 mm\end{tabular} \\
\hline
BSCMR-AS & 599 & Aortic stenosis  & 6 & \begin{tabular}[c]{@{}l@{}}1.5 T, Ingenia, Philips (5.2\%); \\ 1.5 T, Intera, Philips (17.9\%); \\ 1.5 T, Sonata, Siemens (6.2\%);\\ 1.5 T, Aera, Siemens (0.5\%); \\ 1.5 T, Avanto, Siemens (56.6\%);\\ 3 T, Achieva, Philips (0.7\%);\\ 3 T, Skyra, Siemens (3.8\%);\\ 3 T, Verio, Siemens (5.0\%);\\ 3 T, TrioTim, Siemens (4.2\%);\end{tabular} & \begin{tabular}[c]{@{}l@{}}in-plane resolution: \\ 0.78 - 2.3  $mm^2$; \\ slice thickness: \\ 5 - 10 mm\end{tabular} \\ \bottomrule
\end{tabular}%
}
\end{table*}
\textbf{UK Biobank dataset}
The UK Biobank (UKBB) is a large-scale data set that is open to researchers worldwide who wish to conduct a prospective epidemiological study. The UKBB study covers a large population, which consists of over half a million voluntary participants aged between 40 and 69 from across the UK. Besides, the UKBB study performs comprehensive MR imaging for nearly 100,000 participants, including brain, cardiac and whole-body MR imaging. An overview of the cohort characteristics can be found on the UK Biobank’s website~\cite{UKBB_website}. All CMR images we used in this study were collected from one 1.5 Tesla scanner (MAGNETOM Aera, syngo MR D13A, Siemens, Erlangen, Germany). Detailed information about the imaging protocol can be found in~\cite{Petersen2016}. Pixel-wise segmentations of three essential structures (LV, MYO and RV) for both end-diastolic (ED) frames and end-systolic (ES) frames are provided as ground truth~\cite{Petersen2017}. Subjects in this dataset were annotated by a group of eight observers and each subject was annotated only once by one observer. After that, visual quality control was performed on a subset of data to assure acceptable inter-observer agreement.

\textbf{ACDC dataset}
The Automated Cardiac Diagnosis Challenge (ACDC) dataset is part of the MICCAI 2017 benchmark dataset for CMR image segmentation. This set is composed of 100 patients that were evenly divided into 5 classes: dilated cardiomyopathy
(DCM), hypertrophic cardiomyopathy (HCM),
myocardial infarction with altered left ventricular ejection
fraction (MINF), abnormal right ventricle (ARV) and patients
without cardiac disease (NOR). Detailed information about the classification rules and the characteristics of each group can be found in~\cite{Bernard2018} and the ACDC website~\cite{ACDC_website}. All images were collected from one hospital in France. The LV, MYO and RV in this dataset have been manually segmented for both ED frames and ES frames. Images in this dataset were labelled by two cardiologists with more than 10 years of experience~\cite{ACDC_evaluation}.

\textbf{BSCMR-AS dataset}
The British Society of Cardiovascular Magnetic Resonance Aortic Stenosis (BSCMR-AS) dataset from \cite{BSCMRdata} consists of CMR images of 599 patients with severe aortic stenosis (AS), who had been listed for surgery. Images were collected from six hospitals across the UK. Specifically, these images were collected from 9 types of scanner (see Table~\ref{table_for_the_threesdatasets_description}). Although the primary pathology is AS, several other pathologies coexist in these patients (e.g. coronary artery disease, amyloid) and have led to a variety of cardiac phenotypes including left ventricular hypertrophy, left ventricular dilatation and regional infarction \cite{BSCMRdata}. A significant amount of diversity in image appearance and image contrast can be observed in this dataset. Different from the above two data sets, images in this dataset are partially labelled. Only the left ventricle in ED frames and ES frames, as well as the myocardium in ED frames, have been annotated manually. The contours on each slice were refined by an expert. 

In this study, following the same data splitting strategy in \cite{Bai2018}, we split the UKBB dataset into three subsets, containing 3975, 300 and 600 subjects for each set. Specifically, 3975 subjects were used to train the neural network while 300 validation subjects were used for tracking the training progress and avoid over-fitting. The remaining 600 subjects were used for evaluating models' performance. In addition, we directly test this trained network on the other two unseen datasets \emph{without any further re-training or fine-tuning process}. The diversity of pathology observed in the ACDC dataset and the diversity of scanners and cardiac morphologies in the BSCMR-AS make them ideal test sets for evaluating the proposed method's segmentation performance across sites. 

\label{Methods}
\begin{figure*}[!t]
\centering
\includegraphics[width=1.0\textwidth]{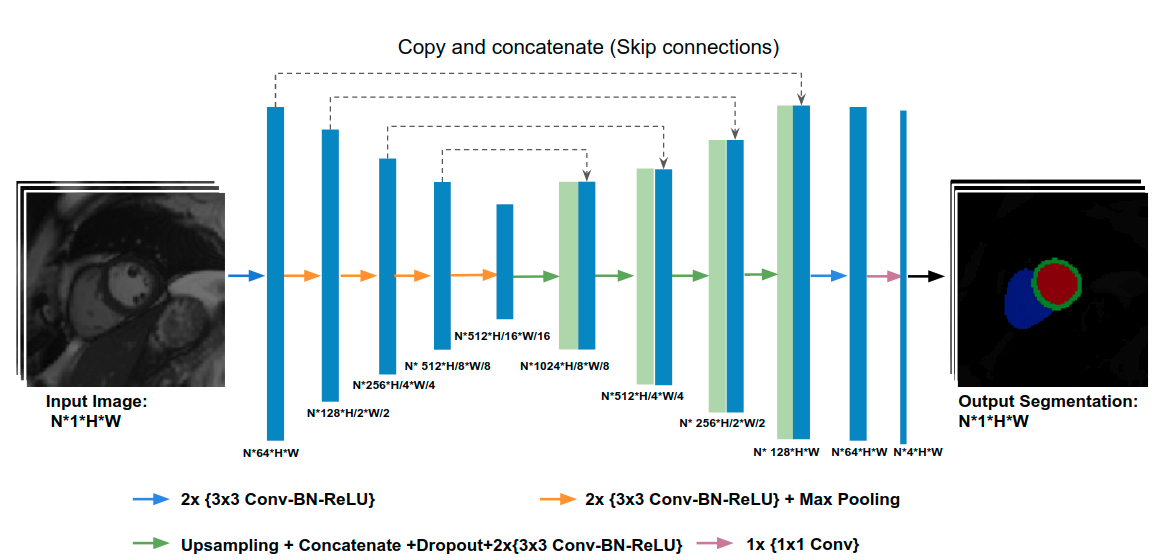}
\caption{\textbf{Overview of the network structure}. Conv: Convolutional layer. BN: Batch normalization. ReLU: Rectified linear unit.
The U-Net takes a batch size of N 2D CMR images as input at each iteration, learning multi-scale features through a series of convolutional layers, max-pooling operations. These features are then combined through upsampling and convolutional layers from coarse to fine scales to generate pixel-wise predictions for the 4 classes (background, LV, MYO, RV) on each slice.}
\label{network_structure}
\end{figure*}
\subsection{Network Architecture}
In this paper, the U-Net architecture~\cite{Ronneberger2015a} is adopted to perform the cardiac multi-structure segmentation task since it is the most successful and commonly used architecture for biomedical segmentation. The structure of our network is illustrated in Fig.~\ref{network_structure}. The network structure is as same as the one proposed in the original paper~\cite{Ronneberger2015a}, except for two main differences. One is that we apply batch normalization (BN)~\cite{ioffe2015batch} after each hidden convolutional layer to stabilise the training. The other difference is that we apply dropout regularization \cite{srivastava2014dropout} after each concatenating operation to avoid over-fitting and encourage generalization.

While both 2D U-Net and 3D U-Net architectures can be used to solve volumetric segmentation tasks~\cite{Isensee2018, Baumgartner2017}, we opt for 2D U-Net for several reasons. Firstly, performing segmentation tasks in a 2D fashion allows the network to work with images even if they have different slice thickness or have severe respiratory motion artefacts between the slices (which is not uncommon). Secondly, 3D networks require much more parameters than 2D networks. Therefore, it is more memory-consuming and time-consuming to train a 3D network than a 2D one. Thirdly, the manual annotation for images in the three datasets were done in 2D (slice-by-slice) rather than 3D. Thus, it is natural to employ a 2D network rather than a 3D network to learn segmentation from those 2D labels.  

\subsection{Training and Testing Pipeline}
\label{SECTION: data preprocessing}
\begin{figure}[!ht]
\includegraphics[width=0.5\textwidth]{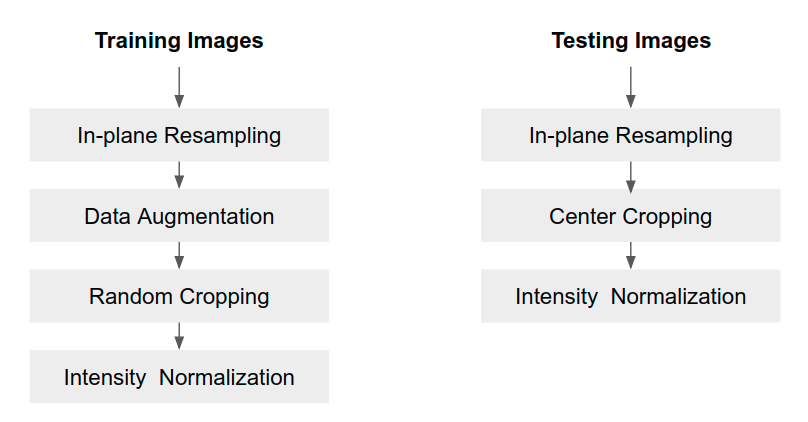}
\caption{\textbf{Image pre-processing during training and testing.}}
\label{data_prep}
\end{figure}

Since training images and testing images in this study were collected from various scanners, it is vital to normalise the input images before feeding them into the network. Fig.~\ref{data_prep} shows an overview of the pipeline for image pre-processing during training and testing. Specifically, we employ image resampling and intensity normalization to normalise images in both the training and testing stages while online data augmentation is applied for improving the model generalization ability during the training process.

\subsubsection{Image Resampling} Since the size of the heart in images with different resolution can vary significantly, it is essential to perform image resampling both in the training and testing phases before cropping such that the proportion of the heart and the background is relatively consistent for segmentation. The pixel spacing in the test image sets ranges from $0.78$ to $2.33~mm$ whereas our training dataset from a single scanner has a uniform pixel spacing of  $1.8~mm$. Therefore, we choose a median value of $1.25~mm$ for resampling. All images are resampled to $1.25\times 1.25~mm$~across short-axis slices. After image resampling, data augmentation is applied to increase the variety of the training set in order to avoid over-fitting and encourage model generalization.

\subsubsection{Data Augmentation} Data augmentation has been widely used when training convolutional neural networks for computer vision tasks on natural images. While different tasks may have different domain-specific augmentation strategies, the common idea is to enhance model's generalization by artificially increasing the variety of training images so that the training set distribution is more close to the test set population in the real world.

In this study, the training dataset is augmented in order to cover a wide range of geometrical variations in terms of the heart pose and size. To achieve this goal, we apply:
\begin{itemize}
    \item random horizontal and vertical flips with a probability of 0.5 to increase the variety of image orientation.
    \item random rotation to increase the diversity of the heart pose. The range of rotation is determined by a hyper-parameter search process. As a result, each time, the angle for augmentation is selected from $[-30, +30]$.
    \item  random image scaling with a scale factor $s$: $s \in [0.7,1.4 ]$ to increase variations of the heart size.
    \item  random image cropping. The random cropping implicitly perform random shifting to augment data context variety without black borders. This operation will also crop images to acceptable sizes required by the network structure. Note that cropping is done after all other image augmentations. Finally, all images are cropped to the same size of $256 \times 256$ before being sent to the network.
\end{itemize} 
We also experimented with contrast augmentation~\cite{Chen2018a} (random gamma correction where the gamma value is randomly chosen from a certain range) to increase image contrast variety, but only minor improvements were found in the experiments. Therefore, it is not included in the pipeline. For each cropped image, intensity normalization with a mean of $0$ and a standard deviation of $1$ is performed.

\subsubsection{Training}
After pre-processing, batches of images are fed to the network for training. To track the training progress, we also use a subset (validation set) from the same dataset to validate the performance of the segmentation and to identify possible over-fitting. Specifically, we apply the same augmentation strategy on the validation set and record the averaged accuracy (mean intersection of union between predicted results and ground truth) on it for each epoch. The model with the highest accuracy is selected as the best model. This selection criterion works as early stopping and has the benefit of allowing the network to explore if there is further opportunity to generalise better before it reaches to the final epoch. 

\subsubsection{Testing}
For testing, 2D images extracted from volume data are first re-sampled and centrally cropped to the same size as the one of the training images. Again, intensity normalization is performed on each image slice which is then passed into the network for inference.  After that, bilinear up-sampling is performed on the outputs of the network to recover the resolution back to the original one. Finally, each pixel of the original image is assigned to the class that has the highest probability among the four classes (background, LV, myocardium, RV). As a result, a final segmentation map for one input image is generated.

\subsection{Implementation Details}
During training, a random batch of 20 2D short-axis slices were fed into the network for each iteration after data pre-processing. The dropout rate for each dropout layer is set to be $0.2$. In every iteration, cross entropy loss was calculated to optimize the network parameters through back-propagation. Specifically, the stochastic gradient descent (SGD) method was used during the optimization, with an initial learning rate of 0.001. The learning rate was decreased by a factor of $0.5$ every 50 epochs. The method was implemented using Python and PyTorch. We trained the U-Net for 1,000 epochs in total which took about 60 hours on one NVIDIA Tesla P40 GPU using our proposed training strategy. During testing, the computation time for segmenting one subject is less than a second.

\subsection{Evaluation Metrics}
The performance of the proposed method was evaluated using the Dice score (3D version) which was also used in the ACDC benchmark study~\cite{Bernard2018} and \cite{Bai2018}. The Dice score evaluates the overlap between automated segmentation $A$ and manual segmentation $B$, which is defined as:
$\textrm{Dice} = \frac{2 |A \cap B|}{|A| + |B|}.$
The value of a Dice score ranges from 0 (no overlap between the predicted segmentation and its ground truth) to 1 (perfect match).

% P-values provided using t-test are two-sided, an alpha level below 0.05 is considered statistically significant.

We also compared the volumetric measures derived from our automatic segmentation results and those from manual ones (see section \ref{clinical measures}), since they are essential for cardiac function assessment. Specifically, for each manual ground truth mask and its corresponding automatic segmentation mask, we calculated the volumes of LV and RV at ED frames and ES frames, as well as the mass of myocardium estimated at  ED frames. The myocardium mass around the LV is estimated by multiplying the LV myocardial volume with a density of 1.05 g/mL. After that, Bland-Altman analysis and correlation analysis for each pair were conducted.  Of note, for Bland-Altman analysis, we removed the outlying mean values that fall outside the range of $1.5 \times \textrm{IQR}$ (interquartile range) in order to avoid the standard deviation of mean difference being biased by extremely large values. These outliers are often associated with poor image quality. As a result, $<3\%$ subjects were removed in each comparison. 

The statistical analysis was performed using python with public packages: \textit{pandas}~\cite{pandas}, \textit{scipy.stats} \cite{scipy}, and \textit{statsmodel}~\cite{statsmodel}.

\section{Results}
\label{results analysis}
We compared the proposed method with the method in our previous work~\cite{Bai2018} in terms of segmentation accuracy across three sets: the UKBB test set, the ACDC set, and the BSCMR-AS set. In~\cite{Bai2018}, a fully convolutional neural network (FCN) was trained using the same UKBB training set and then tested on the same UKBB test set. This method was specifically designed to automatically segment a large scale of scans for the same cohort study with maximum accuracy whereas the proposed method focuses on improving the robustness of the neural network-based segmentation method (using the same UKBB training set as training data) for data from different domains (e.g. non-UKBB data). The comparison results are shown in Table~\ref{TABLE:comparison between bai and ours}. 

\begin{table*}[!h]
\centering
\caption{\textbf{Comparison results of segmentation performance between a baseline method and the proposed method across three test sets}. Both methods were trained using the UKBB training set. The results were evaluated on three sets. Numbers listed in the table are the means and standard deviation of Dice scores.}
\label{TABLE:comparison between bai and ours}
\resizebox{\textwidth}{!}{%
\begin{tabular}{llrrrrrrrr}\toprule
 &  & \multicolumn{3}{c}{\textbf{UKBB Test set (n=600)}} & \multicolumn{3}{c}{\textbf{ACDC set (n=100)}} & \multicolumn{2}{c}{\textbf{BSCMR-AS  set (n=599)}} \\ 
 Method & Training set & \multicolumn{1}{c}{LV} & \multicolumn{1}{c}{MYO} & \multicolumn{1}{c}{RV} & \multicolumn{1}{c}{LV} & \multicolumn{1}{c}{MYO} & \multicolumn{1}{c}{RV} & \multicolumn{1}{c}{LV} & \multicolumn{1}{c}{MYO*} \\ \midrule
Bai, 2018~\cite{Bai2018} & UKBB (n=3975) & 0.94 (0.04) & 0.88 (0.03) & 0.90 (0.05) & 0.81 (0.22) & 0.70 (0.20) & 0.68 (0.31) & 0.82 (0.21) & 0.74 (0.17) \\
Ours & UKBB (n=3975) & 0.94 (0.04) & 0.88 (0.03) & 0.90 (0.05) & 0.90 (0.10) & 0.81 (0.07) & 0.82 (0.13) & 0.89 (0.09) & 0.83 (0.07) \\  \bottomrule
\end{tabular}
}
\begin{tablenotes}
\item *: The myocardium segmentation performance on the BSCMR-AS set was only evaluated on ED frames because of the lack of annotation at ES frames, whereas the performance on the other two datasets was evaluated on both ED and ES frames. For simplicity, Dice scores for the myocardium on the BSCMR-AS in the following tables were calculated in the same way without further illustration.

\end{tablenotes}
\end{table*}

\begin{figure}[!h]
\centering
\includegraphics[width=0.5\textwidth, height=0.4\textheight]{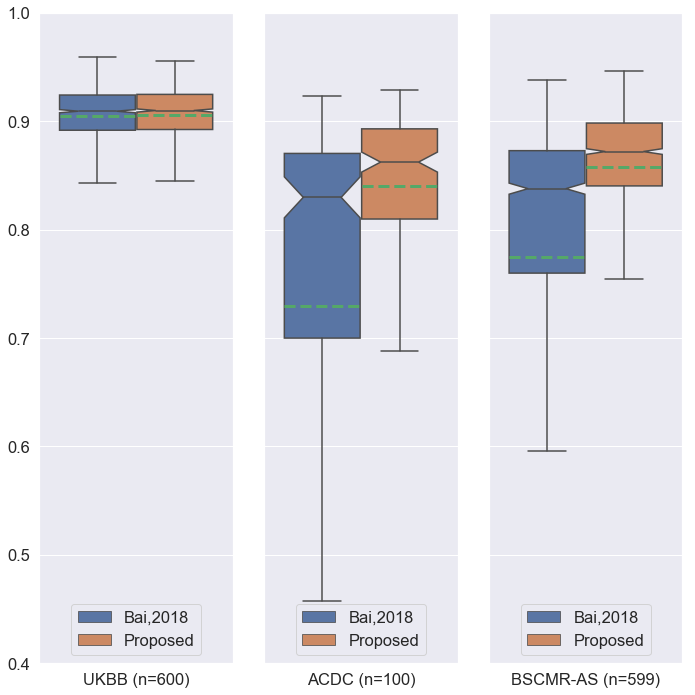}
\caption{\textbf{Boxplots of the average Dice scores between the results of our previous work \cite{Bai2018} and the results of the proposed method on the three datasets.} For simplicity, we calculate the average Dice score over the three structures (LV, MYO, RV) for each image in the three datasets. The boxplots in orange are the results of the proposed method whereas the boxplots in blue are the results of the previous work. The \textcolor{green}{green} dashed line in each boxplot shows the mean value of the Dice scores for the segmentation results on one dataset.}
\label{FCNvsOurs_boxplot}
\end{figure}

While both models achieve very similar Dice scores on the UKBB test set with high accuracy, the proposed method significantly outperforms the approach proposed in \cite{Bai2018} on the two cross-domain datasets: ACDC set and BSCMR-AS set. Compared to the results predicted by \cite{Bai2018} on the ACDC data, the proposed method achieves higher mean Dice scores for all of the three structures: LV  (0.90 vs 0.81), myocardium (0.81 vs 0.70), and RV (0.82 vs 0.68). On the BSCMR-AS dataset, the proposed method also yields higher average Dice scores for the LV cavity (0.89 vs 0.82) and the myocardium (0.83 vs 0.74). Fig.
\ref{FCNvsOurs_boxplot} compares the distributions of Dice scores for the results obtained by the proposed method and the previous work. From the results, the boxplots of the proposed method are shorter than those of the previous method and have higher mean values, which suggests that the proposed method achieves comparatively higher overall segmentation accuracy with lower variance on the three datasets.
% We also conducted a t-test comparing the results obtained from the two methods on each test set, and P-value for each pair is below 0.01 (the last row in Table~\ref{TABLE:comparison between bai and ours}), indicating a significant difference between them. 

% Besides, we tested our model on the BSCMR-AS test subset. The results were closed to a model we trained on BSCMR-AS's training subset (479 subjects) with the same network. We use 'BSCMR-AS model' to refer to this model. Compared to the results predicted by the BSCMR-AS model, our model trained on UKBB got similar averaged dice score on the LV (0.89 vs 0.89) with slightly higher standard deviation (0.08 vs 0.07). For the myocardium segmentation performance, our model obtained a higher averaged dice score on the MYO segmentation (0.84 vs 0.81). Our model outperformed the BSCMR-AS model on the myocardium segmentation, which may because BSCMR-AS model had not seen any ground truth of myocardium on ES frames during training, which limited its learning and inference ability. Our promising results on the ACDC and BSCMR-AS test sets demonstrated that our proposed method greatly closed the domain gap between UKBB and the other two. 

In order to identify what contributes to the improved performance, we further compare the proposed method with~\cite{Bai2018} in terms of methodology. Two main differences are spotted:
\begin{itemize}
    \item \textbf{Network structure and capacity}. Compared to the U-Net we used in this study, FCN in~\cite{Bai2018} has a smaller number of filters at each level. For example, the number of convolutional kernels (filters) in the first layer of FCN is 16 whereas the one in the U-Net is 64. In addition, in the decoder part, FCN directly upsamples the featuremap from each scale to the finest resolution and concatenates all of them, whereas the U-Net adopts a hierarchical structure for feature aggregation. 
    \item \textbf{Training strategy in terms of data normalization and data augmentation}. Compared to the image pre-processing pipeline in the previous work, the proposed pipeline adopts image resampling and random image flip augmentation in addition to the general data augmentation based on affine transformations.
\end{itemize}

In order to study the influence of the network structure as well as the data normalization and augmentation settings on model generalizability, extensive experiments were carried out and the results are shown in the next two sections.
\subsection{The Influence of Network Structure and Capacity}
\label{SECTION: network}
To investigate the influence of network structure on model generalization, we trained three additional networks: 
\begin{itemize}
    \item FCN-16: the FCN network presented in~\cite{Bai2018} which has 16 filters in the first convolutional layer.
    \item FCN-64: a wider version of FCN where the number of filters in each convolutional layer is increased by 4 times.
     \item UNet-16: a smaller version of U-Net where the number of filters in each convolutional layer is reduced by four times. Same as FCN-16, it has 16 filters in the first layer.
\end{itemize}
All of them were trained using the same UKBB training set and with the same training hyperparameters. These networks were then compared to the proposed network (UNet-64).

Table~\ref{TABLE:comparison on the network structure} compares the performances of the four different networks over the three different test sets. It can be seen that while there is no significant performance difference among the four networks on the UKBB test set, small networks: UNet-16 and FCN-16 perform much more poorly than their wider versions: UNet-64 and FCN-64, on the ACDC set (see \textcolor{red}{red numbers} in Table~\ref{TABLE:comparison on the network structure}). This may indicate that in order to accommodate more variety of data augmentation for generalization, the network requires a larger capacity. It is also worth noticing that UNet-64 outperforms FCN-64 on all of the three test sets, while UNet-64 contains fewer parameters than FCN-64. This improvement may result from U-Net's special architecture: skip connections with its step-by-step feature upsampling and aggregation. The results indicate that the network structure and capacity can affect the segmentation model generalizability across datasets. 
% Related literature has shown that this kind of connections can help the network to converge faster and perform better on the segmentation task~\cite{Drozdzal2016}.

\begin{table*}[!h]
\caption{\textbf{Cross-dataset segmentation performances of four different network architectures.} All the networks have been trained using the same UKBB training set with the proposed data normalization and augmentation strategy for \textbf{1,000} epochs. Results listed in the table are the means and standard deviation of the Dice scores evaluated on the three sets. Numbers in \textcolor{red}{red} denote mean Dice scores below 0.70, whereas numbers in the \textbf{bold} font style denote the highest mean Dice scores among the results of the four networks.}
\label{TABLE:comparison on the network structure}
\resizebox{\textwidth}{!}{%
\begin{tabular}{@{}cccccccccc@{}}
\toprule
 &  & \multicolumn{3}{c}{\textbf{UKBB Test set (n=600)}} & \multicolumn{3}{c}{\textbf{ACDC set (n=100)}} & \multicolumn{2}{c}{\textbf{BSCMR-AS  set (n=599)}} \\ \cmidrule(l){3-10} 
\multirow{-2}{*}{\textbf{Network Structure}} & \multirow{-2}{*}{\begin{tabular}[c]{@{}c@{}}\textbf{num of conv} \\ \textbf{weights} (aprox.)\end{tabular}} & LV & MYO & RV & LV & MYO & RV & LV & MYO \\ \midrule
FCN-16 & 0.98 million & 0.92 (0.04) & 0.84(0.04) & 0.88(0.05) & 0.80(0.20) & {\color[HTML]{FE0000} 0.67(0.19)} & {\color[HTML]{FE0000} 0.68(0.27)} & 0.84(0.14) & 0.77(0.11) \\
FCN-64 & 15.6 million & 0.94 (0.04) & 0.87(0.03) & 0.89(0.05) & 0.87(0.12) & 0.78(0.11) & 0.77(0.17) & 0.85(0.12) & 0.79(0.10) \\
UNet-16 & 0.84 million & 0.92 (0.04) & 0.83(0.04) & 0.87(0.05) & 0.87(0.12) & {\color[HTML]{FE0000} 0.66(0.14)} & {\color[HTML]{FE0000} 0.67(0.22)} & 0.85(0.11) & 0.73(0.11) \\
Ours (UNet-64) & 13.4 million & \textbf{0.94 (0.04)} & \textbf{0.88(0.03)} & \textbf{0.90(0.05)} & \textbf{0.90(0.10)} & \textbf{0.81(0.07)} & \textbf{0.82(0.13)} & \textbf{0.88(0.09)} & \textbf{0.83(0.07)} \\ \bottomrule
\end{tabular}
}
% \begin{tablenotes}
% \item FCN-16: The network proposed in \cite{Bai2018}. 16 denotes the number of the convolutional kernels in the first layer. Note that the network had been retrained with the proposed training strategy for comparison.
% \item FCN-64: A wider version of FCN where the number of kernels in each hidden convolutional layer is 4 times larger than those in the FCN-16. 
% \item UNet-16: A smaller U-Net where each level contains the same number of weights as FCN does, such that each level only contains $1/4$ convolutional kernels of the original U-Net. 
% \item UNet-64: Our proposed U-Net which has 64 convolutional kernels in the first layer. 
% \end{tablenotes}

\end{table*}

\subsection{The Influence of Different data normalization and Data Augmentation Techniques}
\label{SECTION: data aug and norm impact}
In this section, we investigate the influence of different data normalization and augmentation techniques on the generalizability of the network, including image resampling (data normalization), scale, flip and rotation augmentation (data augmentation). We focus on these four operations because convolutional neural networks are designed to be translation-equivariant~\cite{Goodfellow2016} but they are not rotation-equivariant, nor scale and flip-equivariant~\cite{bekkers2018roto,rotationSander2015}. This means that if we rotate the input, the networks cannot be guaranteed to produce the same predictions with the corresponding rotation, indicating that they are not robust to geometrical transformations on images. Current methods to improve these networks' ability to deal with rotation/flip/scale variations still heavily rely on data augmentation while intensity-level difference might be addressed by further doing domain adaptation techniques such as style transfer or adaptive batch normalization~\cite{Li2018a}.  

To investigate the influence of these four operations on model generalization, we trained additional three U-Nets using the UKBB training set, each of them was trained with the same settings except that only one operation was removed. To save the computational time for this ablation study, each network was trained for 200 epochs, which still took 10 hours for each network since the training set from the UKBB dataset was considerably large (3,975 subjects). The test results on the UKBB test set, the ACDC dataset, and the BSCMR-AS dataset are shown in Table~\ref{ablation study}. It can be observed that while the results on the test data from the same domain (UKBB) with different settings do not vary much, there are significant differences on the other two test sets, demonstrating the importance of the four data augmentation operations. For example, image resampling increases the averaged Dice score from 0.673 to 0.783 for the RV segmentation on the BSCMR-AS set, whereas augmentation by scaling improves the mean Dice score from 0.596 to 0.750 for the RV on the ACDC set. The best segmentation performance over the three sets is achieved by combining all the four operations. 
\begin{table*}[!ht]
\centering
\caption{\textbf{Cross-dataset segmentation performances of U-Nets with different training configurations.} All experiments were performed with the standard U-Net architecture: UNet-64. Each U-Net was trained using the same UKBB training set for \textbf{200} epochs to save computation. Statistics listed in the table are the means and standard deviation of the Dice scores evaluated on the three sets. Numbers in \textcolor{red}{red} are those mean Dice scores below 0.70.}
\label{ablation study}
\resizebox{\textwidth}{!}{%
\begin{tabular}{@{}cccccccccccc@{}}
\toprule
\multicolumn{4}{c}{\textbf{Configurations}} & \multicolumn{3}{c}{\textbf{UKBB Test set (n=600)}} & \multicolumn{3}{c}{\textbf{ACDC set (n=100)}} & \multicolumn{2}{c}{\textbf{BSCMR-AS set (n=599)}} \\ \midrule
\begin{tabular}[c]{@{}c@{}}Image \\ Resample\end{tabular} & \begin{tabular}[c]{@{}c@{}}Rotation \\ Aug\end{tabular} & \begin{tabular}[c]{@{}c@{}}Flip \\ Aug\end{tabular} & \begin{tabular}[c]{@{}c@{}}Scale \\ Aug\end{tabular} & LV & MYO & RV & LV & MYO & RV & LV & MYO \\ \midrule
\checkmark & \checkmark & \checkmark & \checkmark & 0.923 (0.041) & 0.847 (0.038) & 0.878 (0.048) & 0.873 (0.101) & 0.744 (0.104) & 0.750 (0.187) & 0.851 (0.113) & 0.783 (0.095) \\
\xmark & \checkmark & \checkmark & \checkmark & 0.916 (0.046) & 0.836 (0.041) & 0.864 (0.053) & 0.811 (0.179) & {\color[HTML]{FE0000} 0.614 (0.186)} & {\color[HTML]{FE0000} 0.575 (0.270)} & 0.798 (0.172) & {\color[HTML]{FE0000} 0.673 (0.162)} \\
\checkmark & \xmark & \checkmark & \checkmark & 0.922 (0.042) & 0.848 (0.038) & 0.878 (0.050) & 0.869 (0.117) & 0.733 (0.117) & 0.722 (0.210) & 0.853 (0.118) & 0.784 (0.093) \\
\checkmark & \checkmark & \xmark & \checkmark & 0.924 (0.041) & 0.849 (0.037) & 0.881 (0.049) & 0.858 (0.115) & 0.705 (0.142) & {\color[HTML]{FE0000} 0.681 (0.266)} & 0.862 (0.110) & 0.779 (0.092) \\
\checkmark & \checkmark & \checkmark & \xmark & 0.921 (0.047) & 0.845 (0.039) & 0.876 (0.050) & 0.785 (0.188) & {\color[HTML]{FE0000} 0.640 (0.187)} & {\color[HTML]{FE0000} 0.596 (0.279)} & 0.834 (0.148) & 0.752 (0.125) \\ \bottomrule
\end{tabular}%
}
\end{table*}
These results suggest that increasing variations regarding pixel spacing (image scale augmentation), image orientation (flip augmentation), heart pose (rotation augmentation) as well as data normalization (image resampling) can be beneficial to improve model generalisabilty over unseen cardiac datasets. While one may argue that there is no need to do image resampling if scale augmentation is performed properly during training, we found that image resampling can significantly reduce the complexity of real-world data introduced by heterogeneous image pixel spacings, such that training and testing data are more similar to each other, bringing benefits to both model learning and prediction. In the following sections, we will use `UKBB model' to refer to our best model (the U-Net which was trained using the UKBB training set with our proposed training strategy) for the sake of simplicity.

\subsection{Segmentation Performance on Images from Different Types of Scanners}
\label{SECTION: different scanners}
In this section, UKBB model's segmentation performance is analysed according to different manufacturers (Philips and Siemens) and different magnetic field strengths (1.5 Telsa and 3 Telsa). The results on the two datasets (BSCMR-AS and ACDC) are listed in Table~\ref{Table: segmentation performance across scanners}. For ACDC data, only the results regarding scans imaged using different magnetic strengths are reported since these scans are all from Siemens. Furthermore, results in the ACDC dataset with Dice scores below 0.50 are not taken into account for this evaluation. This is because the number of subjects from a 3T scanner in the ACDC is so small (33 subjects) that the averaged performance can be easily affected given only a few cases with extreme low Dice scores. Here, six subjects were excluded. The final results show that the model trained only using 1.5T Siemens data (UKBB data) could still produce similar segmentation performance on other Siemens and Philips data (top two rows in Table~\ref{Table: segmentation performance across scanners}). Similar results are found on those images acquired from 1.5T scanners and those acquired from 3T scanners (see the bottom four rows in Table~\ref{Table: segmentation performance across scanners}). This indicates that the proposed method has the potential to train a model capable of segmenting images across \textbf{various scanners} even if the training images are only from \textbf{one} scanner. 
\begin{table*}[!ht]
\centering
\caption{\textbf{Segmentation performance of the UKBB model across different scanners}. Tests were performed on the BSCMR-AS dataset and ACDC dataset. This table presents the mean and standard deviation (numbers in the brackets) of the Dice score.}
\label{Table: segmentation performance across scanners}
\resizebox{0.7\textwidth}{!}{%
\begin{tabular}{@{}llllrrr@{}}
\toprule
\textbf{Dataset} & \begin{tabular}[c]{@{}l@{}}\textbf{MRI Scanner} \\ \textbf{Attributes}\end{tabular} & \textbf{Scanners} & \begin{tabular}[c]{@{}l@{}}\# \textbf{of} \\ \textbf{subjects}\end{tabular}& \textbf{LV} & \textbf{MYO} &\textbf{RV}  \\ \midrule
\multirow{4}{*}{BSCMR-AS} & \multirow{2}{*}{Manufactures} & Philips & 142 & 0.89 (0.07) & 0.85 (0.04) & - \\
 &  & Siemens & 457 & 0.88 (0.10) & 0.83 (0.08) & - \\ \cmidrule(l){2-7} 
 & \multirow{2}{*}{Magnetic Field Strengths} & 1.5T & 517 & 0.88 (0.09) & 0.83 (0.09) & - \\
 &  & 3 T & 82 & 0.88 (0.09) & 0.84 (0.09) & - \\ \midrule
\multirow{2}{*}{ACDC} & \multirow{2}{*}{Magnetic Field Strengths} & 1.5T & 65 & 0.89 (0.09) & 0.81 (0.06) & 0.80 (0.09) \\
 &  & 3 T & 29 & 0.91 (0.06) & 0.82 (0.05) & 0.80 (0.08) \\ \bottomrule
\end{tabular}%
}
\end{table*}
\subsection{Segmentation Performance on Images from Different Sites}
\label{SECTION: different sites}
We also evaluate the performance of the UKBB model across seven sites: one from ACDC data, six sites from BSCMR-AS data. Results are shown in Table~\ref{Table: segmentation performance across sites}. From the results, no significant difference is found when evaluating the LV and the myocardium segmentation performances among the seven sites (A-G) while the generalization performance for RV segmentation still needs further investigation when more data with annotated RV becomes available for evaluation.  

\begin{table*}[!h]
\centering
\caption{\textbf{Segmentation performance of the UKBB model across different sites}. This table presents the mean and the standard deviation (numbers in the brackets) of Dice scores for each site.}
\label{Table: segmentation performance across sites}
\resizebox{0.6\textwidth}{!}{%
\begin{tabular}{@{}cccccc@{}}
\toprule
\textbf{Dataset} & \textbf{Site} & \# \textbf{of subjects} & \textbf{LV} & \textbf{MYO} &\textbf{RV} \\ \midrule
% UKBB & \multicolumn{1}{l}{site A} & 600 & \multicolumn{1}{l}{0.94 (0.04)} & \multicolumn{1}{l}{0.88(0.03)} & \multicolumn{1}{l}{0.90(0.05)} \\ \midrule
ACDC & site A & 100 & 0.91 (0.07) & 0.81 (0.08) & 0.82 (0.11) \\ \hline
\multirow{6}{*}{BSCMR-AS} & site B & 28 & 0.88 (0.09) & 0.83 (0.04) & - \\
 & site C & 74 & 0.88 (0.09) & 0.83 (0.04) & - \\
 & site D & 150 & 0.89 (0.07) & 0.85 (0.04) & - \\
 & site E & 122 & 0.86 (0.11) & 0.81 (0.08) & - \\
 & site F & 64 & 0.88 (0.09) & 0.84 (0.08) & - \\
 & site G & 160 & 0.89 (0.09) & 0.85 (0.08) & - \\ \bottomrule
\end{tabular}%
}
\end{table*}

\subsection{Segmentation Performance on Images belonging to Different Pathologies}
\label{SECTION: pathological effects}
\begin{figure*}[!h]
\includegraphics[height=0.3\textheight,width=\textwidth]{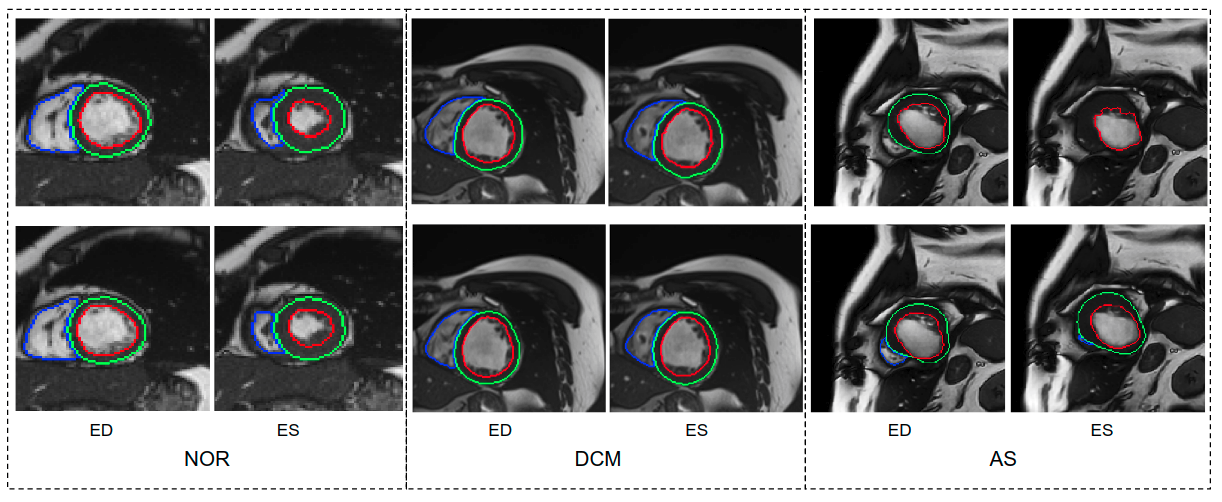}
\caption{\textbf{Visualization of good segmentation examples selected from 3 patient groups.} NOR (without cardiac disease), DCM (dilated cardiomyopathy), AS (aortic stenosis). Each block contains a slice from ED frame and its corresponding ES one for the same subject. Row 1: Ground truth; row 2: predicted results by the UKBB model. This figure shows that the UKBB model produced satisfying segmentation results not only on healthy subjects but also on those DCM and AS cases with abnormal cardiac morphology. The AS example in this figure is a patient with aortic stenosis who previously had a myocardial infarction.}
\label{Fig.good results from pathological groups}
\end{figure*}

We further compare the segmentation performance of the proposed method on five groups of pathological data to the group of normal subjects (NOR), see Table~\ref{Table: segmentation performance across pathology}. Surprisingly, the UKBB model achieves satisfying segmentation accuracy over the healthy group as well as DCM images and those images diagnosed with AS, indicating the model is capable of segmenting not only those with normal cardiac structures but also some abnormal cases with the cardiac morphological variations in those HCM images and AS images, see Fig.~\ref{Fig.good results from pathological groups}. 

\begin{table*}[!h]
\caption{\textbf{Segmentation performance of the UKBB model across the five groups of pathological cases and normal cases (NOR)}. This table presents the mean and standard deviation of the Dice score. \textcolor{red}{Red} numbers are those mean Dice scores below 0.80.}
\label{Table: segmentation performance across pathology}
\centering
\resizebox{0.6\textwidth}{!}{%
\begin{tabular}{cccccc}
\toprule
\textbf{Dataset} & \textbf{Group} & \# \textbf{of subjects} & \textbf{LV} & \textbf{MYO} &\textbf{RV}  \\ \midrule
 & NOR & 20 & 0.91 (0.05) & 0.83 (0.04) & 0.85 (0.14) \\
 & DCM & 20 & 0.94 (0.04) & 0.81 (0.05) & 0.82 (0.11) \\
 & HCM & 20 & 0.84 (0.12) & 0.84 (0.03) & 0.84 (0.08) \\
 & MINF & 20 & 0.92 (0.05) & 0.81 (0.04) & {\color[HTML]{FE0000} 0.78 (0.13)} \\
\multirow{-5}{*}{ACDC} & ARV & 20 & 0.86 (0.13) & {\color[HTML]{FE0000} 0.74 (0.11)} & {\color[HTML]{FE0000} 0.79 (0.16)} \\ \hline
BSCMR-AS & AS & 599 & 0.88 (0.09) & 0.83 (0.07) & - \\ \bottomrule
\end{tabular}%
}
\end{table*}
\begin{figure}[!ht]
\includegraphics[width=0.5\textwidth]{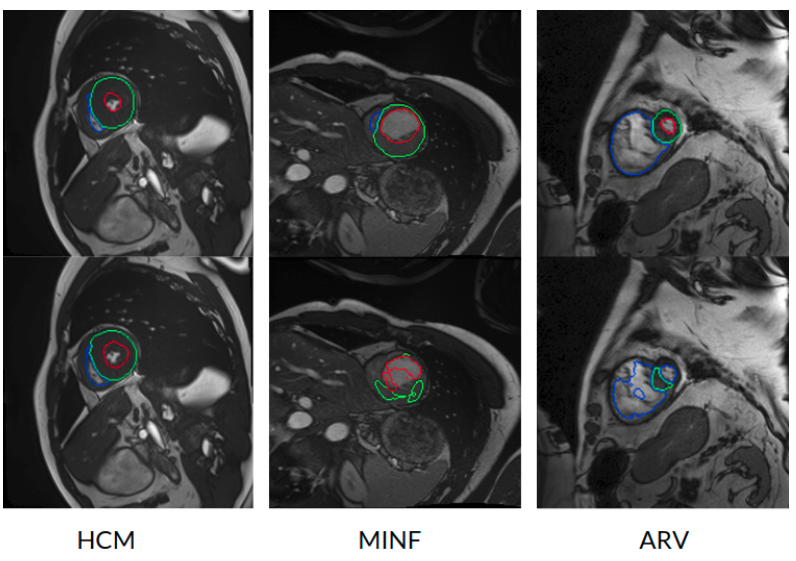}
\caption{\textbf{Examples of the worst cases that have pathological deformations.} Row 1: Ground truth; row 2: predicted results by the UKBB model. HCM: hypertrophic cardiomyopathy; MINF: myocardial infarction with altered left ventricular ejection fraction; ARV: abnormal right ventricle. Column 1 shows that the UKBB model underestimates the myocardium in patients with HCM. Column 2 shows that the model struggles to predict the cardiac structure when certain sections of the myocardium are extremely thin. Column 3 shows a failure case where an extremely large right ventricle is shown in the image.}
\label{Fig.bad results from pathological groups}
\end{figure}

However, the model fails to segment some of the other pathological images, especially those in the HCM, MINF, and ARV pathology groups where lower Dice scores are observed. For example, the mean Dice score for LV segmentation on  HCM images is the lowest (0.84). Fig.~\ref{Fig.bad results from pathological groups} demonstrates some of the worst cases produced by our method. The first column in Fig.~\ref{Fig.bad results from pathological groups}, shows a failure case where the UKBB model underestimated the myocardium and overestimated the LV when a thickened myocardial wall is present in a patient with HCM. Also, the model struggles to segment cardiac structure on a patient with MINF which contains the abnormal myocardial wall with non-uniform thickness (the second column in Fig.~\ref{Fig.bad results from pathological groups}). Compared to images in the other four groups, images from patients with ARV seem to be more difficult for the model to segment as the model not only achieves a low mean Dice score on the RV (0.79) but also a low averaged value on the myocardium (0.74).

One possible reason for these unsatisfactory segmentation results might be the lack of pathological data in the current training set. In fact, the UKBB data only consists of a small amount of subjects with self-reported cardiovascular diseases, and the majority of the data are healthy subjects in middle and later life~\cite{Petersen2017,fry2017comparison,Bai2018}. This indicates that the network may not be able to `learn' the range of those pathologies that are seen in everyday clinical practice, especially those abnormalities which are not currently reprepresented in the UKBB dataset.

\begin{figure}[!h]
\includegraphics[width=0.5\textwidth]{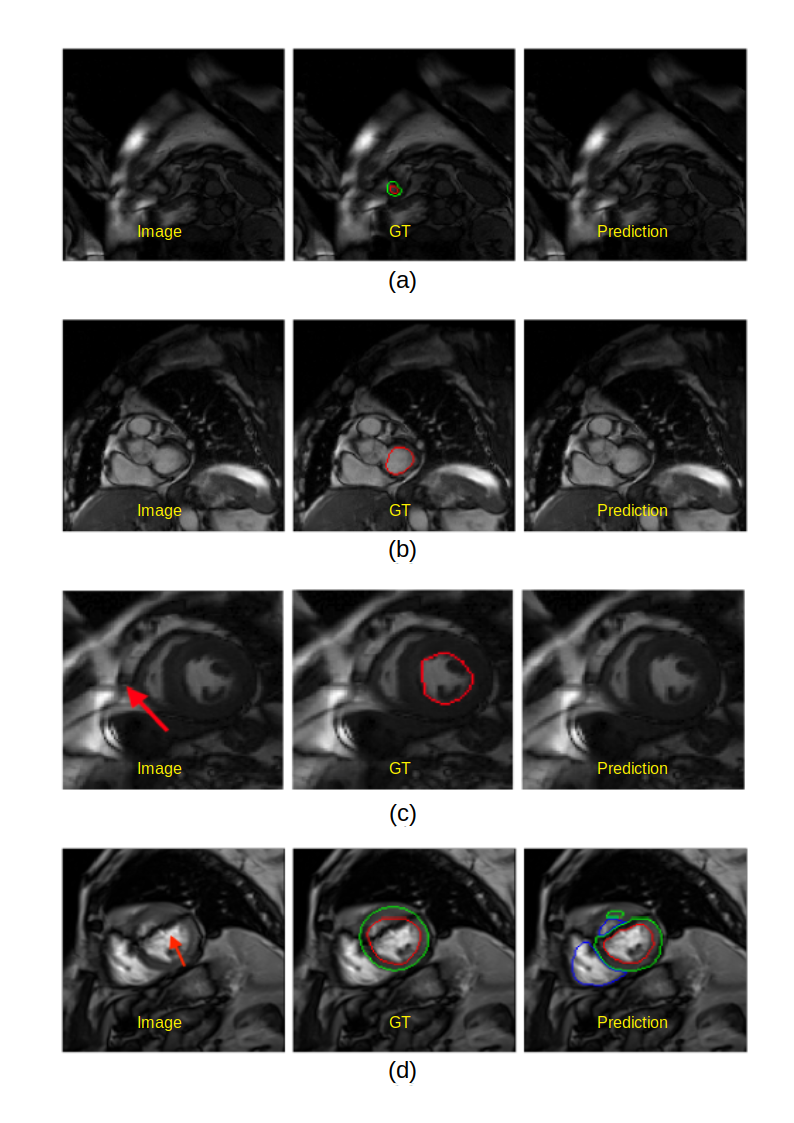}
\caption{\textbf{Examples of worst segmentation results found on challenging slices}. Left: Image, middle: ground truth (GT), right: prediction from the UKBB model.
(a) Failure to predict LV when the apical slice has a very small LV.
(b) LV segmentation missing on the basal slice (ES frame) . This sample is from the BSCMR-AS dataset where only the LV endocardial annotation is avaiable.
(c) Failure to recognize the LV due to a stripe of high-intensity noise around the cardiac chambers in this 1.5T image. This sample is an ES frame image from the BSCMR-AS dataset.
(d) Failure to estimate the LV structure when unexpected strong dark artefacts disrupt the shape of the LV in this 3T image. Note that this image is an ED frame image from the BSCMR-AS dataset where RV was not annotated by experts.}
\label{Bad results due to image quality and position}
\end{figure}

\textbf{Failure Mode Analysis}
We also visually inspected the images where the UKBB model produces poor segmentation masks. In general, there are two main failure modes we identified, apart from the failure found on the abnormal pathological cases which we have discussed above:
\begin{itemize}
    \item \textbf{Apical and basal slices}. These slices are more error-prone than mid-ventricle slices, which has also been reported by~\cite{Bernard2018}.  Segmenting these slices is difficult because apical slices have extremely tiny objects which can be hard to locate and segment (see Fig.~\ref{Bad results due to image quality and position} (a)) whereas basal slices with complex structures increase the difficulty of identifying the contour of the LV (see Fig.~\ref{Bad results due to image quality and position} (b)). 
    \item \textbf{Low image quality}. Images with poor quality are found both in 1.5T and 3T images (see Fig.~\ref{Bad results due to image quality and position} (c) and (d)). As reported in~\cite{Rajiah, Alfudhili2016}, 1.5T images are more likely to have low image contrast than 3T images due to the low signal-to-noise (SNR) limits, whereas 3T images can have more severe imaging artefact issues than 1.5T images. These artefacts and noise can greatly affect the segmentation performance.
\end{itemize}

\subsection{Statistical Analysis on Clinical Parameters}
\label{clinical measures}
\begin{figure*}[!ht]
\centering
\includegraphics[width=0.85\textwidth, height=0.85\textheight]{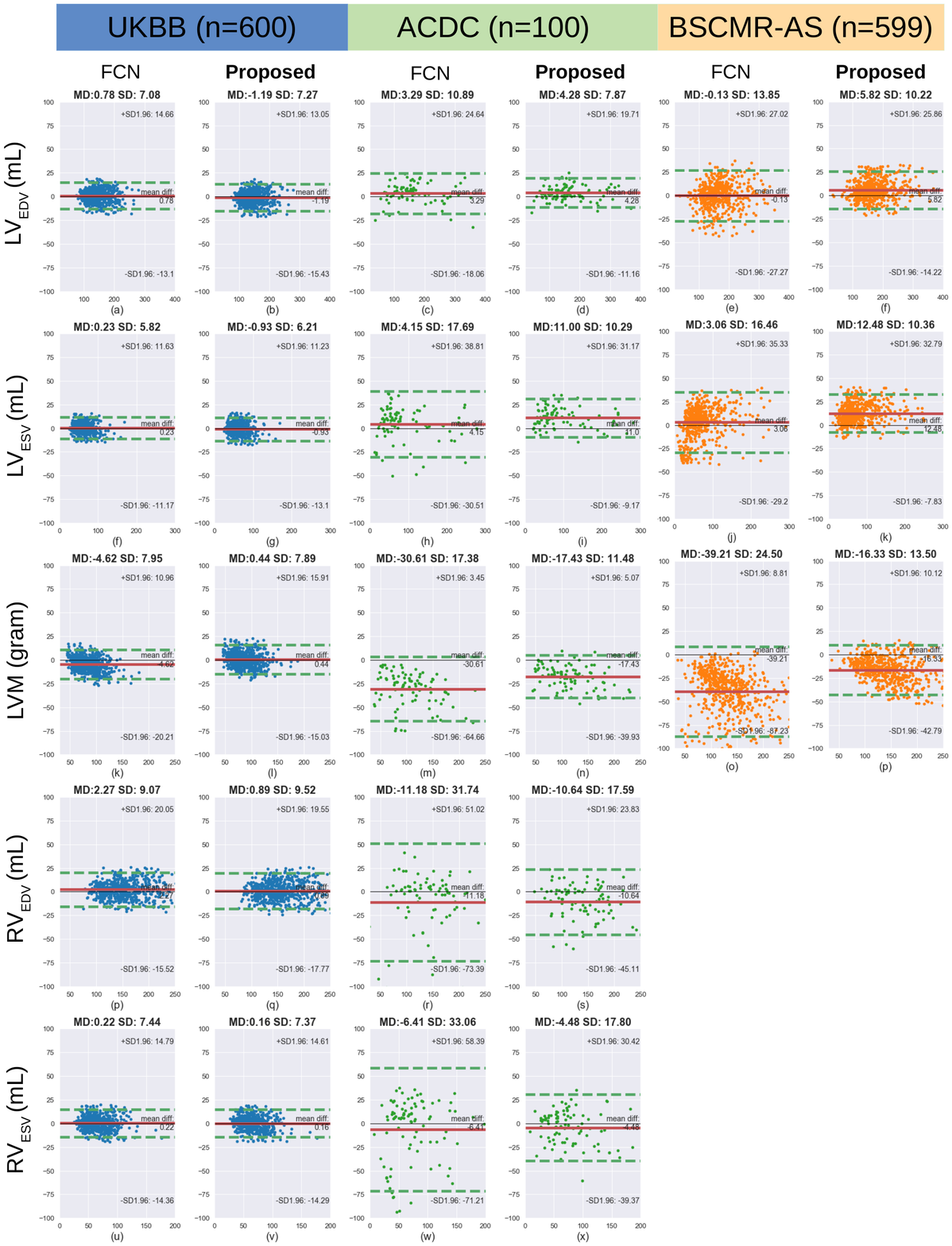}
\caption{\textbf{Agreement of clinical measurement from automatic and manual segmentations.} Bland Altman plots (automatic - manual) are shown regarding the three sets. In each Bland-Altman plot, the x-axis denotes the average of two measurements whereas the y-axis denotes the difference between them. The solid line in \textcolor{red}{red} denotes the mean difference (bias) and the two dashed lines in \textcolor{green}{green} denote $ \pm 1.96$ standard deviations from the mean. The title of each plot shows the mean difference (MD) and its standard deviation (SD) for each pair of measurements. FCN: the automatic method in our previous work~\cite{Bai2018}, LV/RV: left/right ventricle, EDV/ESV: end-diastolic/systolic volume, LVM: left ventricular mass.}
\label{bland_altman plots}
\end{figure*}
We further compare the proposed automatic method with manual approach on five clinical parameters, including the end-diastolic volume of LV ($LV_{EDV}$),  the end-systolic volume of LV (${{LV}}_{{ESV}}$), the left ventricular mass ($LVM$), the end-diastolic volume of right ventricle (${RV}_{{EDV}}$), and the end-systolic volume of RV ($RV_{ESV}$). 

Figure \ref{bland_altman plots} shows the Bland-Altman plots for the five clinical parameters on the three datasets. The Bland-Altman plot is commonly used for analysing agreement and bias between two measurements. Here, each column shows the comparison results between automated measurements and manual measurements for one particular parameter, including the mean differences (MD) with corresponding standard deviation (SD) and the limits of agreement (LOA). In addition, we also conducted the Bland-Altman analysis for the automatic method (FCN) in our previous work \cite{Bai2018}, for ease of comparison.

From the first two columns in the Fig.~\ref{bland_altman plots}, one can see that both FCN and the proposed method achieve excellent agreements with human observers on the UKBB dataset, indicating both of them can be used interchangeably with manual measurements. For the other two datasets, by contrast, the proposed method achieves much better agreement than FCN, as the LOA between the proposed method and manual results is narrower. For example, for $LVM$ on the ACDC dataset, the LOA between the proposed method and the manual approach is from 5.07 to -39.93 (MD =-17.43) while the LOA between the FCN and the manual method is from 3.45 to -64.66 (MD = -30.61), see Fig.~\ref{bland_altman plots} (n) and Fig.~\ref{bland_altman plots} (m), respectively. 

Finally, we calculate the Spearmanr's rank correlation coefficients ($r^2$) of the five clinical parameters derived from the automatic segmentation using the proposed method and the manual segmentation, which are reported in Table~\ref{correlation}. From the results, it can be observed that the clinical measurements based on the LV segmentation and the myocardium segmentation derived by our automatic model are highly positively correlated with the manual analysis ($\geq0.91$), although the RV correlation coefficients on the ACDC dataset are relatively lower. 
\begin{table*}[!h]
\centering
\caption{\textbf{Spearman's rank correlation coefficients of clinical parameters derived from the automatic measurements and the manual measurements on the three sets.} All segmentations are produced by the U-Net trained with the UKBB training set.}
\label{correlation}
\begin{threeparttable}[]
\begin{tabular}{@{}cccccccc@{}}
\toprule
\textbf{Comparison} & \textbf{Test set} & \begin{tabular}[c]{@{}c@{}}$\mathbf{LV_{EDV}}$\\
\end{tabular} & \begin{tabular}[c]{@{}c@{}}$\mathbf{LV_{ESV}}$\\ \end{tabular} & \begin{tabular}[c]{@{}c@{}}$\mathbf{LVM}$\\ \end{tabular}  & \begin{tabular}[c]{@{}c@{}}$\mathbf{RV_{EDV}}$\\ \end{tabular} & \begin{tabular}[c]{@{}c@{}}$\mathbf{RV_{ESV}}$\\ \end{tabular} \\ \midrule
Automatic vs Manual & UKBB  (n=600) & 0.97 & 0.91 & 0.93  & 0.96 & 0.91 \\
Automatic vs Manual & ACDC (n=100) & 0.97 & 0.94 & 0.96  & 0.79 & 0.83 \\
Automatic vs Manual & BSCMR-AS (n=599) & 0.94 & 0.92 & 0.92 & -  & - \\ \bottomrule
\end{tabular}%
\begin{tablenotes}
\item [Note:] Each coefficient reported in this table has a P-value below 0.0001.
\end{tablenotes}
\end{threeparttable}
\end{table*}
\section{Discussion}
\label{limitation}
In this paper, we developed a general training/testing pipeline based on data normalization and augmentation for improving the generalizability of neural network-based CMR
image segmentation methods. We also highlighted the importance of the network structure and capacity (section~\ref{SECTION: network}) as well as the data normalisation and augmentation strategies (section~\ref{SECTION: data aug and norm impact}) for model generalizability. Extensive experiments on multiple test sets were conducted to validate the effectiveness of the proposed method. The proposed method achieves promising results on a large number of test images from various scanners and sites even though the training set is from one scanner, one site (section \ref{SECTION: different scanners},~\ref{SECTION: different sites}). Besides, the network is capable of segmenting healthy subjects as well as a group of pathological cases from multiple sources although it had only been trained with a \emph{small} portion of pathological cases. 

The limitation of the current method (the UKBB model) is that it still tends to underestimate the myocardium especially when the size of the myocardium becomes larger (see points in the right part of Fig.~\ref{bland_altman plots} (p)). Again, we conclude this limitation is mainly due to the lack of pathological cases in the training set. 

Besides, we found that the difference (bias) between the automatic measurements and the manual measurements in the cross-domain test sets: ACDC and BSCMR-AS, are more significant than the difference in the intra-domain set: UKBB test set.  The larger bias may be caused by not only those challenging pathological cases we have discussed above, but also inter-observer bias and the inconsistent labelling protocols used in the three datasets. The evident inter-observer variability when delineating myocardial boundaries on apical and basal slices in a single dataset has been reported in \cite{Suinesiaputra2014}. In this study, however, there are three datasets which were labelled by three \emph{different} groups of observers. Each group followed an independent labelling protocol. As a result, significant variations of RV labels and MYO labels on the basal planes among the three datasets are found. This inter-dataset inconsistency of the RV labels on basal planes has been reported in \cite{zheng20183}. The mismatch of RV labels can partially account for the negative MD values for the RV measurements in the ACDC dataset (see Fig.~\ref{bland_altman plots} (s)). The differences in the labelling protocols together with inter-observer variability in different datasets pose challenges to evaluate the model generalizability across domains accurately.

In the future, we will focus on improving the segmentation performance of the neural network by increasing the diversity of the training data in terms of pathology. A promising way of doing it, instead of collecting more labelled data, is to synthesize pathological cases by transforming existing healthy subjects with pathological deformations. A pioneering work~\cite{J_Krebs_2019} in this direction has successfully transported pathological deformations from certain pathological subjects (i.e. HCM, DCM) to healthy subjects, which can help to increase the number of pathological cases. This deformation-based data augmentation can be easily integrated in the proposed training pipeline without significant modifications.

\section{Conclusion}
\label{conclusion}
In this paper, we proposed a general training/testing pipeline for neural network-based cardiac segmentation methods and revealed that a proper design of data normalization and augmentation, as well as the design of network, play essential roles in improving its generalization ability across images from various domains. We have shown that a neural network (U-Net) trained with CMR images from a \textbf{single} scanner has the potential to produce competitive segmentation results on \textbf{multi-scanner} data across domains. Besides, experimental results have shown that the network is capable of segmenting healthy subjects as well as a group of pathological cases from multiple sources although it had only been trained with the UK Biobank data which has only a \emph{small} portion of pathological cases. Although it might still have the limitations in segmenting images with low quality and some images with significant pathological deformations, higher segmentation accuracy for these subjects could be further achieved by increasing the diversity of training data regarding image quality and the pathology in the future.

\section*{Abbreviations}
CNN: convolutional neural network; MR: magnetic resonance; CMR: cardiac magnetic resonance; UDDA: unsupervised deep domain adaptation; LV: left ventricle; MYO:myocardium; RV: right ventricle; ED: end-diastole; ES: end-systole; FCN: fully convolutional network; GPU: graphics processing unit; UKBB: UK Biobank; ACDC: Automatic Cardiac Diagnosis Challenge; MICCAI: International Conference on Medical Image Computing and Computer-Assisted Intervention; DCM: dilated cardiomyopathy; HCM: hypertrophic cardiomyopathy; MINF: myocardial infarction with altered left ventricular ejection fraction; ARV: abnormal right ventricle; NOR: without cardiac disease; AS: aortic stenosis; BSCMR: the British Society of Cardiovascular Magnetic Resonance; SGD: stochastic gradient descent; SNR: signal-to-noise; EDV: end-diastolic volume; ESV: end-systolic volume; LVM: left ventricular mass; MD: mean difference; SD:standard deviation of mean difference; LOA: limits of agreement. 

\section*{Ethics approval and consent to participate}
UK Biobank has approval from the North West Research Ethics Committee (REC reference: 11/NW/0382). The ACDC data has been fully anonymised and handled within the regulations set by the local ethical committee of the Hospital of Dijon (France). The BSCMR-AS data has approval from the UK National Research Ethics Service (REC reference:13/NW/0832), and has been conformed to the principles of the Declaration of Helsinki. All patients included in the BSCMR-AS study gave written informed consent.

\section*{Consent for publication}
Not applicable.

\section*{Competing interests}
The authors declare that they have no competing interests.

\section*{Availability of data and material}
Imaging data and manual annotations of the UKBB dataset were provided by the UK Biobank Resource under Application Number 2964. Researchers can apply to use the UK Biobank data resource for health-related research in the public interest~\cite{UKBiobankReg2017}.
The ACDC data is open to the public and can be downloaded from its website~\url{https://acdc.creatis.insa-lyon.fr/#challenges} after registration. The BSCMR-AS dataset is available upon reasonable request.
The code for training and testing the segmentation network will be available at \url{https://github.com/cherise215/CardiacMRSegmentation}. The code is used for data pre-processing, data augmentation, and the segmentation network training and testing.

\section*{Funding}
This work is supported by the SmartHeart EPSRC Programme Grant (EP/P001009/1). C.M. is supported directly and indirectly by the University College London Hospitals, NIHR Biomedical Research Centre and Biomedical Research Unit at Barts Hospital, respectively. S.N., E.L., S.K.P. are supported by the Oxford NIHR Biomedical Research Centre. S.E.P., S.K.P. and S.N. acknowledge the British Heart Foundation (BHF) for funding the manual analysis to create a cardiovascular magnetic resonance imaging reference standard for the UK Biobank imaging resource in 5000 CMR scans (PG/14/89/31194).

\section*{Authors' contributions}
C.C., W.B. and D.R. conceived and designed the study; Rh.H.D., A.N.B, C.M. and J.C.M provided support on clinical aspects and they also provided the BSCMR-AS data resource to be used for testing; N.A., A.M.L., M.M.S., K.F., J.M.P., S.E.P., E.L., S.K.P., S.N. provided the UKBB data resource to be used for training and testing and support on clinical aspects; C.C. designed the method, performed data analysis and wrote the manuscript. All authors read and approved the manuscript.

\section*{Acknowledgment}
This research has been conducted mainly using the UK Biobank Resource under Application Number 40119 and 2964. The authors wish to thank all UK Biobank, ACDC, and BSCMR-AS participants and staff.

\bibliographystyle{unsrt}
\bibliography{mybib}
\end{document}